# Dependence of critical parameters of 2D Ising model on lattice size

**B.V. Kryzhanovsky[1], M.Yu. Malsagov[1], I.M. Karandashev[1,2]**

[1]Scientific Research Institute for System Analysis, Russian Academy of Sciences,
[2]Peoples Friendship University of Russia (RUDN University)
malsagov@niisi.ras.ru, karandashev@niisi.ras.ru, kryzhanov@gmail.com

**Abstract.** For the 2D Ising model, we analyzed dependences of thermodynamic characteristics on number of spins by means of computer simulations. We compared experimental data obtained using the Fisher-Kasteleyn algorithm on a square lattice with $N = l \times l$ spins and the asymptotic Onsager solution ($N \to \infty$). We derived empirical expressions for critical parameters as functions of $N$ and generalized the Onsager solution on the case of a finite-size lattice. Our analytical expressions for the free energy and its derivatives (the internal energy, the energy dispersion and the heat capacity) describe accurately the results of computer simulations. We showed that when $N$ increased the heat capacity in the critical point increased as $\ln N$. We specified restrictions on the accuracy of the critical temperature due to finite size of our system. Also in the finite-dimensional case, we obtained expressions describing temperature dependences of the magnetization and the correlation length. They are in a good qualitative agreement with the results of computer simulations by means of the dynamic Metropolis Monte Carlo method.

## 1. Introduction

Calculation of a partition function is the central problem of statistical physics. However, they succeeded in solving this problem only for a few models described in classical monographs [1, 2]. The most known is the Onsager solution for a two-dimensional square lattice [3] obtained long ago. Until now, there are no analytical solutions for more complex models including models where a magnetic field is taken into account. To describe these models they use approximate methods such as the Bragg-Williams approximation, the approximation of the Curie-Weiss molecular field, the Bethe-Peierls approximation [4, 5], and so on. In spite of its difficulty, the transfer matrix method showed itself very useful when analyzing properties of the Ising model [6, 7]. In addition, a lot of to analytical and numerical methods showed themselves useful when analyzing the Ising model. They are the finite chain extrapolations [8], the high-temperature series, the two-time Green function [10], the Suzuki-Trotter transformation [11], the Monte Carlo method [12, 13], the Bethe approximation [14], a combinatorial approach [15], low-temperature power series [16, 17], and the n-vicinity method [18, 19]. Different authors used the statistical physics methods for analyzing properties of associative memory [20]-[23] and for developing new ways for finite-size neural networks learning [24]-[26].

Recently researchers made significant progress in development new numerical methods for calculation of critical exponents [27]-[30] and energy spectrums of spin systems [31]. In all these studies, the Monte Carlo method is the basic algorithm allowing one to obtain approximate estimates. In the same time, there appeared algorithms for accurate calculation of the free energy of a planar spin system. To approximate $N \to \infty$ in the best way physical results corresponding to the limit the researchers perform calculations with the most possible number of spins $N$. However, the resources of numerical methods are limited and always there is an open question whether the dimension of the problem is sufficiently large.

In the present paper, to perform our computer simulations we used the Fisher-Kasteleyn algorithm [32-33]. This algorithm allowed us to calculate in a polynomial time the free energy of a spin system for an arbitrary planar graph with arbitrary connections. This algorithm is exact because it calculates a partition function by means of calculation of a determinant of a matrix constructed with regard of the model in question. In [34] one can find more about this algorithm. In our calculations, we used the algorithm implementation proposed in [35]. Its results coincide with the results of [34] but it works much faster.

In this paper, our purpose was to investigate the dependencies of the system parameters on the number of spins $N$ and to obtain analytical expressions suit for finite values of $N$. The obtained results allowed us to estimate how large $N$ had to be for the numerical calculations to describe satisfactory properties of asymptotically large ($N \to \infty$) physical models. The present paper is an extended version of our preceding paper [39] and this is why we make good use of the results of this paper. Here we refine some results of [39] and add computer simulations for small linear lattices sizes ($l = 5,...,40$). In addition, in the finite-dimensional case we obtain and analyze new expressions for temperature dependences of the magnetization and the correlation length of spins that are in good agreement with the results of computer simulations by means of the dynamic Metropolis Monte Carlo method.

## 2. Basic expressions

Let us define a spin system described by Hamiltonian

$$E = -\frac{1}{2N} \sum_{i,j=1}^{N} J_{ij} s_i s_j, \quad s_i = \pm 1. \tag{1}$$

We are interested in its free energy

$$f = -\frac{\ln Z}{N}, \tag{2}$$

where we define the partition function $Z = \sum_S \exp[-N\beta E(S)]$ as the sum over all possible configurations $S$, and $\beta$ is the inverse temperature. When we know the free energy, we can calculate the basic measurable parameters of the system. They are

$$U = \frac{\partial f}{\partial \beta}, \quad \sigma^2 = -\frac{\partial^2 f}{\partial \beta^2}, \quad C = -\beta^2 \frac{\partial^2 f}{\partial \beta^2} \tag{3}$$

where the internal energy $U = \overline{E}$ is the ensemble average at given $\beta$, $\sigma^2 = \overline{E^2} - \overline{E}^2$ is the energy dispersion, and $C = \beta^2 \sigma^2$ is the heat capacity.

Let us examine a square lattice with spins in the lattice sites. Each spin interacts only with its four nearest neighbors. We restrict ourselves to the case when interaction with nearest neighbors is defined by the one constant $J_{ij} = J$. In what follows we show that generalization to the case of different interaction constants $J \neq J'$ is straightforward.

The Onsager solution [3] corresponds to the limit $N \to \infty$ and periodic boundary conditions. For simplicity we set $J = 1$, then the expressions for the free energy, the internal energy and the heat capacity take the forms:

$$f = -\frac{\ln 2}{2} - \ln(\cosh 2\beta) - \frac{1}{2\pi} \int_0^\pi \ln\left(1 + \sqrt{1 - k^2 \cos^2\theta}\right) d\theta$$

$$U = -\coth 2\beta \cdot \left[1 + \frac{2}{\pi} \mathbf{E}_1(k)(2\tanh^2 2\beta - 1)\right] \tag{4}$$

$$C = \frac{4\beta^2}{\pi \tanh^2 2\beta} \cdot \left\{\mathbf{E}_1(k) - \mathbf{E}_2(k) - (1 - \tanh^2 2\beta)\left[\frac{\pi}{2} + (2\tanh^2 2\beta - 1)\mathbf{E}_1(k)\right]\right\}$$

where

$$k = \frac{2\sinh 2\beta}{\cosh^2 2\beta} \tag{5}$$

and $\mathbf{E}_1 = \mathbf{E}_1(k)$ and $\mathbf{E}_2 = \mathbf{E}_2(k)$ are the complete elliptic integrals of the first and the second kinds, defined as:

$$\mathbf{E}_1(x) = \int_0^{\pi/2} (1 - x^2 \sin^2\varphi)^{-1/2} d\varphi, \quad \mathbf{E}_2(x) = \int_0^{\pi/2} (1 - x^2 \sin^2\varphi)^{1/2} d\varphi. \tag{6}$$

It is known, that in the framework of the Onsager solution the heat capacity diverges logarithmically ($C \to \infty$) when $\beta \to \beta_{ONS}$, where the critical value of inverse temperature and critical values of free energy and internal energy are determined from the condition $k = 1$ in the form:

$$\beta_{ONS} = \frac{1}{2}\ln(1+\sqrt{2})$$

$$f_{ONS} = -\frac{\ln 2}{2} - \frac{2G}{\pi} \approx -0.9297 \tag{7}$$

$$U_{ONS} = -\sqrt{2}$$

where $G \approx 0.915966$ is Catalan's constant.

In the following sections, we describe the results of our computer simulations. We used the Fisher-Kasteleyn algorithm for lattices of different sizes and compared our numerical results with the analytic Onsager solution. In particular, we analyzed the behavior of the heat capacity near the critical point $\beta_c = \beta_c(N)$. We also examined how $\beta_c$ changes when the dimension of the system increased. Here we present expressions allowing us to generalize the Onsager solution on the case of systems of finite sizes.

### 3. Experimental conditions

Using the Fisher-Kasteleyn algorithm we succeeded in analysis of the behavior of the free energy and its derivatives (the internal energy, the energy dispersion and the heat capacity) for several lattices of different sizes. Linear sizes of our lattices vary from $l=5$ to $l=10^3$. Note, here we limited ourselves to square lattices; however, the algorithm can be used not for them only.

We emphasize that the algorithm we used is suitable for planar lattices only. This means that we examine the case of the free boundary conditions because a lattice with the periodic boundary conditions is not a planar graph. Then the energy of the ground state has the form

$$E_0 = -2\left(1 - \frac{1}{\sqrt{N}}\right). \tag{8}$$

Here we restricted ourselves to linear sizes $l \leq 10^3$ and varied the temperature inside the interval $\beta \in [0,1]$ because when $\beta > 1$ the value of the free energy practically does not differ from its asymptotic value $f \approx \beta E_0$. For each $N$ inside the specified interval $\beta \in [0,1]$, we calculated the function $f = f(\beta)$ and its derivatives with the step size equal to $d\beta = 10^{-4}$. Outside this interval, we calculated the free energy at $10^2$ control points only. In our simulation, to obtain the first and the second derivatives of the function $f = f(\beta)$ we calculated the values of the free energy at three points equispaced at $d\beta = 10^{-5}$ and then used the method of finite differences in the double-precision format. Although the algorithm allowed us to calculate the value of $f$ for a lattice of an arbitrary size, for large linear sizes ($l > 400$) the accuracy was not enough and this is the reason why we calculated the first derivative to an accuracy of $10^{-10}$ and the second derivative to an accuracy of $10^{-5}$ only.

**4. Results of simulations**

In our simulations, we calculated the free energy $f = f(\beta)$ and its derivatives. As expected, we obtained that the peak of the heat capacity $C = C(\beta)$ is shifted to the right with respect to the peak of the energy dispersion $\sigma = \sigma(\beta)$. Let the critical inverse temperature $\beta_c$ be the coordinate of the peak of the heat capacity and let the corresponding critical values be $f_c = f(\beta_c)$, $U_c = U(\beta_c)$, $\sigma_c = \sigma(\beta_c)$ and $C_c = C(\beta_c)$. The coordinate of the peak of the energy dispersion defines the second critical point $\beta_c^*$. The critical values corresponding to $\beta_c^*$ we denote as $f_c^* = f(\beta_c^*)$, $U_c^* = U(\beta_c^*)$, $\sigma_c^* = \sigma(\beta_c^*)$, and $C_c^* = C(\beta_c^*)$. In Table 1 we present the critical values $\beta_c$, $f_c$, $U_c$, $\sigma_c$, and $C_c$ as well as the critical values $\beta_c^*$, $f_c^*$, $U_c^*$, $\sigma_c^*$, and $C_c^*$.

**Table 1.** Critical values corresponding to peak of heat capacity/energy dispersion

| $l$ | $\beta_c / \beta_c^*$ | $|f_c| / |f_c^*|$ | $|U_c| / |U_c^*|$ | $\sigma_c / \sigma_c^*$ | $C_c / C_c^*$ |
|---|---|---|---|---|---|
| 4 | 0.6034 / 0.4554 | 1.0138 / 0.8667 | 1.1436 / 0.8317 | 1.7876 / 2.2743 | 0.6508 / 0.4716 |
| 5 | 0.5680 / 0.4620 | 1.0015 / 0.8875 | 1.2053 / 0.9373 | 2.2337 / 2.6787 | 0.7206 / 0.5717 |
| 8 | 0.5176 / 0.4656 | 0.9786 / 0.9160 | 1.2925 / 1.1118 | 3.2537 / 3.5885 | 0.8716 / 0.7779 |
| 10 | 0.5016 / 0.4644 | 0.9700 / 0.9236 | 1.3203 / 1.1733 | 3.7641 / 4.0476 | 0.9470 / 0.8729 |
| 16 | 0.4776 / 0.46 | 0.9553 / 0.9322 | 1.3572 / 1.2691 | 4.8899 / 5.0668 | 1.1153 / 1.0721 |
| 20 | 0.4700 / 0.4578 | 0.9504 / 0.9340 | 1.3697 / 1.3023 | 5.4323 / 5.5726 | 1.2000 / 1.1679 |
| 25 | 0.4642 / 0.4556 | 0.9467 / 0.9350 | 1.3808 / 1.3287 | 5.9751 / 6.0902 | 1.2875 / 1.2641 |
| 30 | 0.4602 / 0.4534 | 0.9440 / 0.9346 | 1.3871 / 1.3430 | 6.4269 / 6.5208 | 1.3611 / 1.3405 |
| 40 | 0.4550 / 0.4510 | 0.9401 / 0.9346 | 1.3930 / 1.3643 | 7.1519 / 7.2118 | 1.4806 / 1.4669 |
| 50 | 0.4522 / 0.4494 | 0.9382 / 0.9343 | 1.3985 / 1.3768 | 7.7072 / 7.7559 | 1.5760 / 1.5664 |
| 64 | 0.4494 / 0.4476 | 0.9360 / 0.9335 | 1.4003 / 1.3853 | 8.3372 / 8.3635 | 1.6837 / 1.6756 |
| 100 | 0.4462 / 0.4454 | 0.9337 / 0.9326 | 1.4054 / 1.3978 | 9.4658 / 9.4753 | 1.8845 / 1.8797 |
| 200 | 0.4434 / 0.4432 | 0.9317/ 0.9314 | 1.4098 / 1.4075 | 11.2172 / 11.2236 | 2.2053 / 2.2046 |
| 300 | 0.4422 | 0.9306 | 1.4078 | 12.2506 | 2.3955 |
| 400 | 0.4418 | 0.9304 | 1.4091 | 12.9974 | 2.5369 |
| 500 | 0.4418 | 0.9305 | 1.4131 | 13.5659 | 2.6479 |
| 600 | 0.4414 | 0.9301 | 1.4104 | 14.0814 | 2.7435 |
| 700 | 0.4414 | 0.9302 | 1.4124 | 14.5477 | 2.8344 |
| 800 | 0.4414 | 0.9302 | 1.4139 | 14.8560 | 2.8945 |
| 900 | 0.4414 | 0.9303 | 1.4152 | 14.9787 | 2.9184 |
| 1000 | 0.4412 | 0.9301 | 1.4132 | 15.1428 | 2.9477 |

The results of simulations and the analysis of the obtained experimental data we show below in Figs. 1-4. As we see from Fig. 1a, our numerically calculated values for the free energy and for the internal energy approach the Onsager solution when the size of the lattice increases. We show only the curves for not so large lattices whose linear sizes are $l = 25, 50, 100$. When $l > 100$ the curves are too close to the Onsager solution to be clearly seen.

For large $\beta$ mainly the states with the smallest energies contribute to the partition function. Consequently, the free energy for large $\beta$ decreases linearly (see Fig. 1a). According to Eq. (8) when $\beta$ is large asymptotic behavior of the free energy is as follows

$$f \approx -2\beta\left(1 - \frac{1}{\sqrt{N}}\right) \tag{9}$$

Because of the term $\sim 1/\sqrt{N}$, when linear sizes are not too large the curves differ from the Onsager solution.

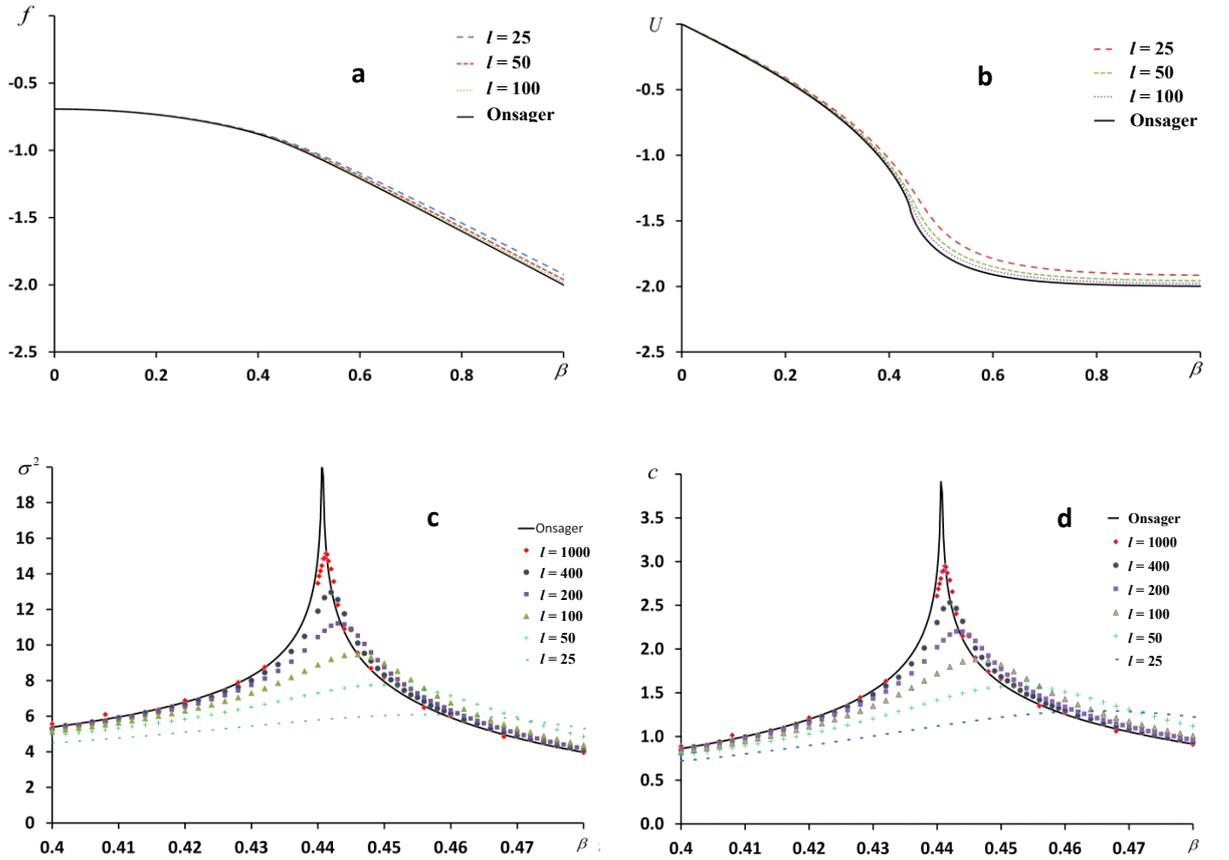

**Fig. 1.** Computer simulations: (a) free energy and (b) internal energy as functions of $\beta$ for not too large linear sizes of lattice ($l = 25, 50, 100$) and asymptotic Onsager solution ($l \to \infty$). Dependences of (c) dispersion and (d) specific heat for different sizes $l$ (marks); solid lines are asymptotic Onsager solution ($l \to \infty$).

In the case of lattices of finite sizes, the second derivative of the free energy shows the difference between the Onsager solution and the function $f(\beta)$ most clearly. According to the Onsager solution the heat capacity diverges logarithmically when $\beta \to \beta_{ONS}$. From Fig. 2 it is evident that for lattices of finite sizes this is not the case. However, the heat capacity $C = C(\beta)$ has a peak and the larger the size of the lattice the sharper the peak.

Analyzing in detail the behavior of the heat capacity near the peak, we found that, first, the height of the peak depends logarithmically on the size of the lattice. Second, the coordinate of the peak is slightly shifted to the right with respect to $\beta_{ONS}$ but it tends to this value when the size of the lattice increases.

In Fig. 1d the solid line is the Onsager solution for the case $N \to \infty$. Note, when $N$ decreases the peaks shift to the right (in the direction of larger values of $\beta$). In [40] the same computer simulations was performed for the two-dimensional Ising model but contrary to the present results, the peaks of the heat capacity shifted to the left. The point is that we examine two-dimensional lattices with the free boundary conditions however in [40] was used the periodic boundary conditions.

Analysis of data presented in Table 1 shows that the expressions (10) approximate correctly the dependences of the coordinate of the peak of the heat capacity (the critical value $\beta_c$) and also of the critical values of the free energy, internal energy and the heat capacity on $N$:

$$\beta_c = \beta_{ONS} \cdot \left(1 + \frac{5}{4l} + \frac{1}{l^2}\right)$$

$$f_c = f_{ONS} \cdot \left(1 + \frac{0.4572}{l}\right)$$

$$U_c = -\sqrt{2} \cdot \left(1 - \frac{1}{2l}\right)$$

$$C_c = \frac{4\beta_c^2}{\pi}\left(\ln N - 1.7808\right)$$

(10)

When approximating the dependence $\beta_c = \beta_c(N)$ by the aid of Eq. (10) the relative error is comparatively small. Its maximal value ~0.3% corresponds to $l = 25$; when $l$ increases the error decreases rapidly up to the value 0.01% for $l = 10^3$. The relative errors of approximation of the values of $U_c$ and $C_c$ are less than 0.4% and 0.8%, respectively. As it follows from Eqs. (10), the critical value of the dispersion defined as $\sigma_c^2 = C_c / \beta_c^2$ depends linearly on $\ln N$. In Fig. 2 we clearly see that Eqs. (10) are in good agreement with the experimental data. Note only that fluctuations seen for $l > 400$ are due to lack of accuracy of our algorithm in the double-precision format to calculate correctly the second derivative in the 4-th decimal place. We note that the correction of order $\sim O(l^{-2})$ in (10) is important only for $l < 20$; for $l \geq 20$ this correction does not influence the accuracy of the approximation.

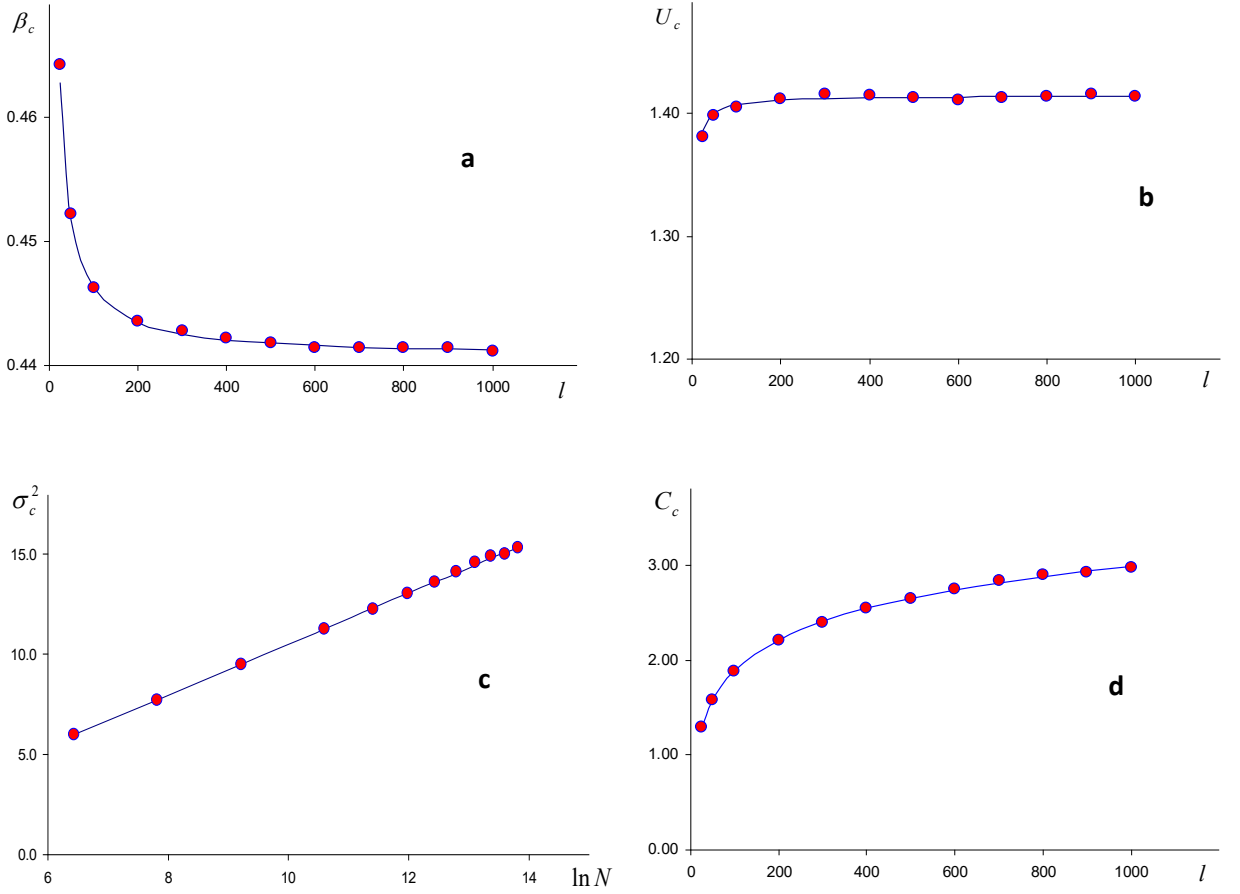

**Fig. 2.** Dependences of critical values of (a) temperature, (b) internal energy, (c) energy dispersion, and (d) heat capacity on the linear size *l*: circles are experimental data, solid lines correspond equations (10). In graph (c) we use logarithmic scale to accentuate linear dependence of critical value of dispersion on $\ln N$.

From data of Table 1, it follows that Eqs. (11) approximate correctly the coordinate of the peak of the energy dispersion (the critical value $\beta_c^*$) and the values of the free energy, internal energy and the heat capacity corresponding to this peak:

$$\beta_c^* = \beta_{ONS} \cdot \left(1 + \frac{1}{l} - \frac{4.2}{l^2}\right)$$

$$f_c^* = f_{ONS} \cdot \left(1 + \frac{0.3629}{l} - \frac{5.7541}{l^2} + \frac{13.03}{l^3}\right)$$

$$U_c^* = -\sqrt{2} \cdot \left(1 - \frac{1}{l}\right) \qquad (11)$$

$$C_c^* = 1.197\, \beta_{ONS}^2 \cdot (\ln N - 1)$$

These expressions are in good agreement with the experimental data. The maximal value of the relative error corresponds to $l=25$ and it is equal to 0.6%, 2.1%, and 1.2% for $\beta_c^*$, $U_c^*$, and $C_c^*$, respectively. When $l$ increases the values of the relative errors decrease rapidly and for $l=10^3$ they are 0.02%, 0.03%, and 0.08%, respectively. In Eq. (11), the function $f_c^* = f_c^*(l)$ is more complex than the analogous function in Eq. (10). In the same time, the accuracy of the approximation is very high: the value of the related error is less than 0.02% even when $l=5$. Note, in Eq. (11) we need to take into account the terms of the order of $O(l^{-2})$ only if $l<40$; when the linear size of the lattice is larger ($l \geq 40$) they are practically do not affect the accuracy of the approximation. Fig. 3 where we show the function $f_c^* = f_c^*(l)$ illustrates this statement. We see that when $l=40$ the terms of the order of $O(l^{-2})$ lead to inflections of the curves; if $l>40$ these terms do not actually influence the shapes of the curves.

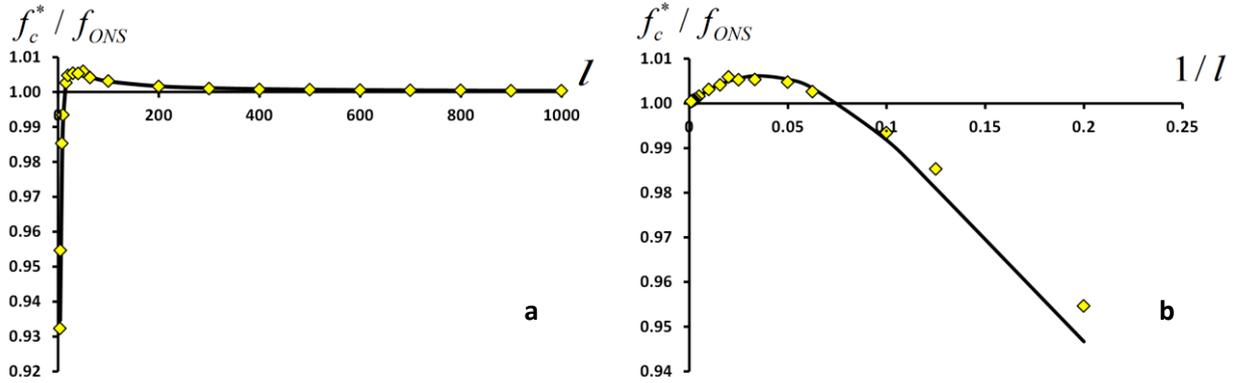

**Fig.3.** (a) $f_c^*$ vs $l$; (b) $f_c^*$ vs $l^{-1}$. The values of free energy $f_c^*$ are normed to Onsager solution. Circles are data from Table 1 ($l=5,...,1000$) and solid line corresponds to Eq. (11).

### 5. Generalized Onsager solution

Our analysis showed that it is possible to derive analytical expressions describing correctly the results of computer simulations and the above-found asymptotic expressions. With this in mind we substitute

$$2\beta \to z = \frac{2\beta}{1+\Delta}, \quad k \to \kappa = \frac{2\sinh z}{(1+\delta)\cosh^2 z}. \tag{12}$$

in Eq. (4), where $\Delta \ll 1$ and $\delta \ll 1$. The purpose of the first substitution is to take boundary conditions into account. As we see from Eq. (8) the energies of the ground state for the free boundary conditions and the periodic boundary conditions coincide up to a constant that differs from one by a value of the order of $O(1/l)$. The second substitution eliminates the divergence at $k \to 1$ ($\kappa < 1$ when $\delta > 0$).

Substituting the expressions (12) in Eq. (4) we obtain the expression for the free energy and after that from Eqs. (3), we calculate the internal energy and the heat capacity applicable for systems of finite sizes:

$$f(\beta) = -\frac{\ln 2}{2} - \ln(\cosh z) - \frac{1}{2\pi} \int_0^\pi \ln\left(1 + \sqrt{1 - \kappa^2 \cos^2 \theta}\right) d\theta$$

$$U = -\frac{1}{1+\Delta}\left\{2\tanh z + \frac{\sinh^2 z - 1}{\sinh z \cdot \cosh z}\left[\frac{2}{\pi} \mathbf{E}_1(\kappa) - 1\right]\right\} \quad (13)$$

$$C = \frac{z^2}{\pi \tanh^2 z} \cdot \left\{a_1(\mathbf{E}_1(\kappa) - \mathbf{E}_2(\kappa)) - (1 - \tanh^2 z)\left[\frac{\pi}{2} + (2a_2 \tanh^2 z - 1)\mathbf{E}_1(\kappa)\right]\right\}$$

where

$$a_1 = p(1+\delta)^2, \quad a_2 = 2p - 1, \quad p = \frac{(1 - \sinh^2 z)^2}{(1+\delta)^2 \cosh^4 z - 4\sinh^2 z}, \quad (14)$$

and $\mathbf{E}_1(\kappa)$ and $\mathbf{E}_2(\kappa)$ are the complete elliptic integrals of the first and the second kinds, defined in (6). As we might expect in the limit $N \to \infty$ from Eq. (14) it follows that $p \to 1$, $a_{1,2} \to 1$ and consequently Eqs. (13) go over to the well-known Onsager expressions (4).

Comparing Eqs. (13) with the experimental curves $f = f(\beta)$, $U = U(\beta)$, $\sigma = \sigma(\beta)$, and $C = C(\beta)$ we see that the best coincidence takes place when the fitting parameters $\Delta$ and $\delta$ are

$$\Delta = \frac{5}{4\sqrt{N}}, \quad \delta = \frac{\pi^2}{N} \quad (15)$$

### 6. Validation of generalized Onsager solution

Analyses of Eqs. (13) shows that the corrections for finite sizes of lattices lead to insignificant changes in the behavior of the free energy and the internal energy. However, these corrections eliminate the logarithmical divergence in the expression for the heat capacity at the critical point. Indeed from Eqs. (13) it follows that the heat capacity reaches its maximum when $\sinh z = 1 + o(N^{-1})$. Omitting here the correction of the order of $o(N^{-1})$ we obtain the expression for the critical temperature:

$$\beta_c = \beta_{ONS}(1+\Delta) \quad (16)$$

Now substituting $\Delta$ from Eq. (15) in Eq. (16), we see that this expression is just the same as the empirical formula (10). Expanding the function $C(\beta)$ in the vicinity of the critical point $\beta_c$ and omitting the terms polynomial in $\beta - \beta_c$ we obtain

$$C(\beta) \approx \frac{4\beta_c^2}{\pi} \left\{ 3\ln 2 - \frac{\pi}{2} - \ln\left[ 4J^2(\beta - \beta_c)^2 + \frac{\pi^2}{N} \right] \right\} \tag{17}$$

From Eq. (17) we obtain the following expression for the critical value of the heat capacity

$$C_c = \frac{4\beta_c^2}{\pi}\left( \ln N + 3\ln 2 - 2\ln \pi - \frac{\pi}{2} \right), \tag{18}$$

Since $2\ln \pi + \pi/2 - 3\ln 2 \approx 1.7808$ the obtained Eq. (18) coincides with the empirical formula (10). In the same way Eqs. (13) confirm our empirical formulae (10) – (11).

Even when the size of the lattice is comparatively small, Eqs. (13) approximate correctly the results of our computer simulations. To illustrate this in Fig. 4 we show the graphs of the energy dispersion, the heat capacity and the internal energy when $N = 25 \times 25$. As we see these figures demonstrate good agreement of the theory with the experiment. The agreement improves significantly, when the size of the lattice increases.

We can also verify the expressions (13) analyzing the spectral function $\Psi(E) = N^{-1} \ln D(E)$ where $D(E)$ is the divergence of the energy level $E$. As it was shown in [37,40] the following equations

$$E = \frac{df(x)}{dx}, \quad \Psi(E) = xE - f(x), \tag{19}$$

define implicitly the spectral function. In Eq. (19) $f(x)$ is the function $f(\beta)$ from Eq. (13) where $\beta$ is replaced by $x$. We introduce an arbitrary variable x to point out that the spectral function is not related to the temperature. Since for each x from Eq. (19) we can find the energy $E$ and the corresponding value of the function $\Psi(E)$, changing $x$ from 0 to $\infty$ we obtain the whole energy spectrum. For verification, we used the results of very complex simulations [38]. The maximal size of their 2D model was $N = 16 \times 16$. In Fig. 4d we compare the results of this experiment with the function $\Psi(E)$ calculated using Eq. (13). We see that even for such small size of the lattice our theory describes well the experimental data.

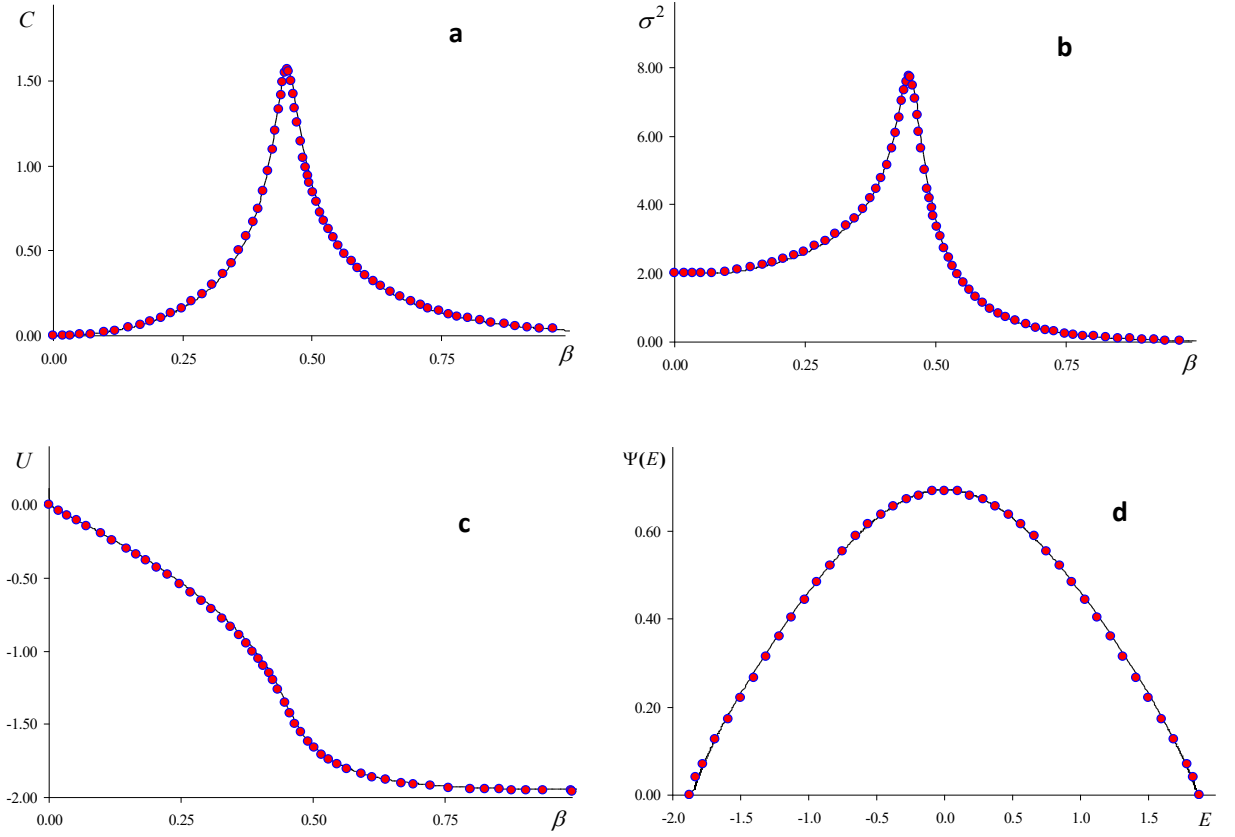

**Fig.4.** Dependences of (a) heat capacity, (a) energy dispersion, and (c) internal energy on $\beta$ for $N = 25 \times 25$: solid lines correspond to Eq. (13), circles are results of simulations. Graph (d) shows spectral functions for lattice with $N = 16 \times 16$ spins: solid line corresponds to Eq. (19) and circles are experimental results [38].

Our equations (13) allow us to examine the dependence of the correlation length $\xi$ on $N$. In the limit $N \to \infty$, in accordance with the well-known expression from [1] this dependence is $\xi = -1/2 \ln \eta$, where $\eta = k/(1 + \sqrt{1 - k^2})$. To use this expression for a lattice of a finite size we make the substitution (12) $k \to \kappa$. Then we obtain the correlation length in the form

$$\bar{\xi} = -\frac{1}{2 \ln \bar{\eta}}, \quad \bar{\eta} = \frac{\kappa}{1 + \sqrt{1 - \kappa^2}} \tag{20}$$

At the critical point $\beta = \beta_c$, the value $\kappa = \kappa(z)$ reaches its maximum equal to $\kappa_{max} = 1/(1+\delta)$ and the correlation length is minimal and equal to

$$\bar{\xi}_{mzx} = \frac{l}{2\pi\sqrt{2}}. \tag{21}$$

Approximately this value one order of magnitude less than the linear size of the lattice $l$.

The correlation length $r_d = \exp(-d/\bar{\xi})$ characterize the correlation between spins at a distance $d$. Taking into account Eq. (20), we obtain the expression for $r_d$ in the form:

$$r_d = \bar{\eta}^{2d} \tag{22}$$

Figures 5-6 shows the experimental values of the correlation between four nearest spins ($d = 1,2,3,4$), as well as the theoretical curves constructed from this formula. The results are obtained by the Metropolis Monte Carlo method.

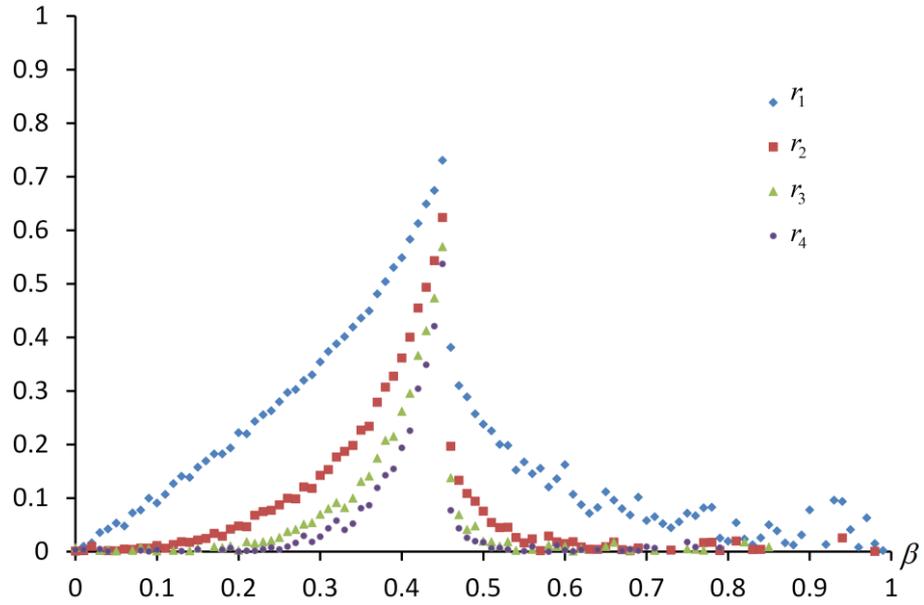

**Fig. 5.** Correlation $r_d$ as a function of temperature: lattice size $l = 50$ and $d = 1,2,3,4$, counting from the center of the lattice.

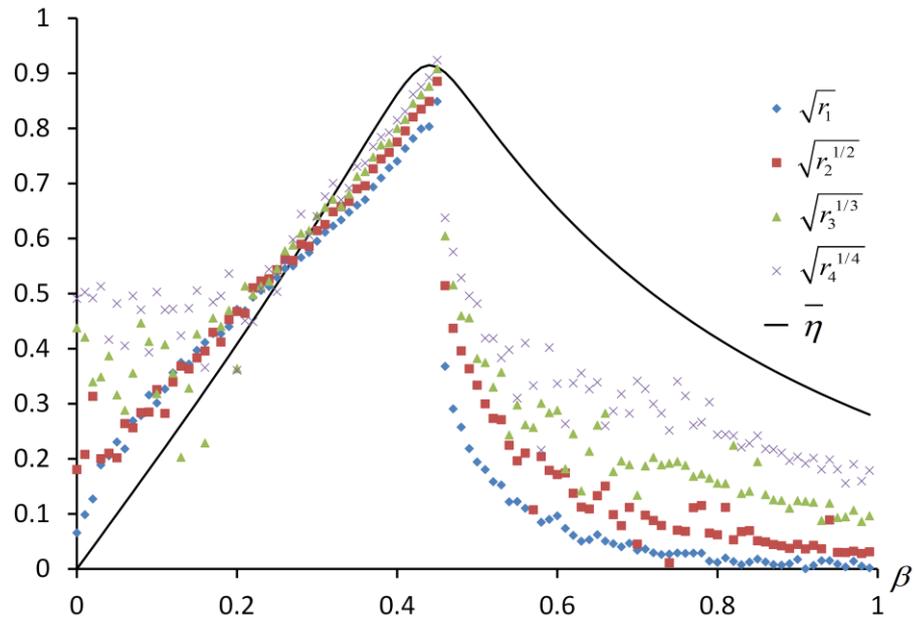

**Fig. 6.** Graphs $\sqrt[2d]{r_d}$ vs $\beta$ for $d = 1,2,3,4$ for square lattice of size $l = 50$. The solid line shows the theoretical curve $\bar{\eta}(\beta)$.

In the same way, we substitute Eqs. (12) in the expression for the spontaneous magnetization obtained by Yang [36]. Then in our case the spontaneous magnetization is

$$M_0 = \left[1 - \left(\frac{\bar{\eta}}{\bar{\eta}_0}\right)^2\right]^{1/8} \qquad (23)$$

where $\bar{\eta}_0 \approx 1/(1+\sqrt{2\delta})$ is the value of the function $\bar{\eta} = \bar{\eta}(\kappa)$ for $\beta = \beta_c$ ($M_0 = 0$ when $\beta < \beta_c$). It must be emphasized, that when the size of the system is finite the interpretation of the expression (23) is completely different than in the case $N \to \infty$. As it was pointed out in [1], in the absence of magnetic field mean values of magnetizations of finite systems are equal to zero. The reason is that for each configuration with $s_i = +1$ there is the equally probable configuration with $s_i = -1$. Consequently Eq. (23) states that at different instants of time the magnetization obtained in the framework of simulations takes on any value between $M_0$ and $-M_0$.

To verify the obtained theoretical values of the magnetization for each temperature we examined experimentally the distribution of the magnetization by means of the Metropolis Monte Carlo method. Since the mean value of the magnetization is always equal to zero we checked the value of magnetization at the peak of distribution only. When $\beta$ is not so large the distribution has one peak at $M = 0$. When $\beta$ is close to the critical value or larger the peak splits. In Fig. 7, we show the coordinates of the magnetization peak as function of the temperature. As we see, they are in good agreement with the obtained theoretical curve.

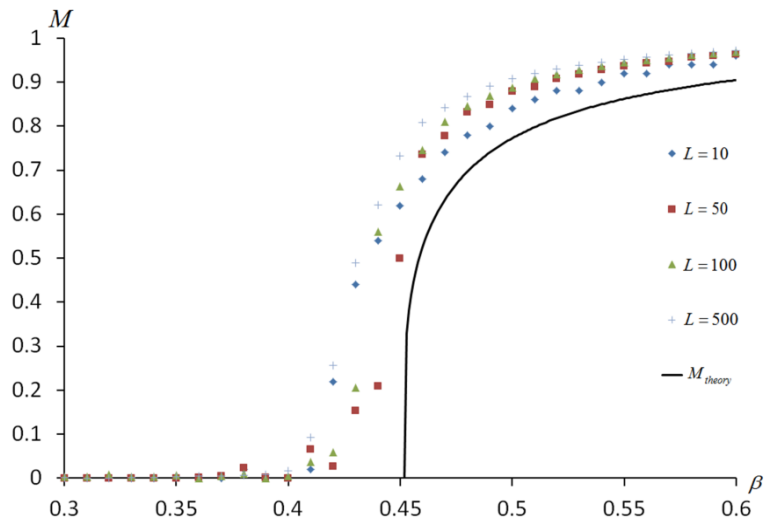

**Fig. 7.** Dependence of the magnetization $M$ on the reciprocal temperature $\beta$ for different values $l = 10...500$. The solid line is the theoretical curve (23) for $l = 50$ ($M = 0$ when $\beta < \beta_c$).

## 7. Discussion and conclusions

Basing on our experimental data, we obtained simple expressions (10) and (11) allowing us to estimate the critical values that are in good agreement with experiments. If we choose the fitting parameters in the form (15) the error of approximation in the limit $\beta \to \infty$ is rather small and we can describe the coordinates and heights of peaks almost precisely. Of course, one may try to define the fitting parameters more accurately and then write down the estimates for critical values to higher accuracy. However, it was not our goal since we seek for the simplest expressions providing the high accuracy when $l \geq 40$. The objective of our work was to find how the critical values behave themselves as functions of $N$.

We introduced two fitting parameters (15) and this allows us to derive analytic expressions (13) generalizing the Onsager solution to the case of lattices of finite sizes. With great accuracy, these expressions describe the behavior of spin systems even when sizes of their lattices are small ($N \sim 25 \times 25$). When $N \geq 50 \times 50$ the differences between our analytical solution and the experimental data become less than experimental errors. Moreover, introduction of these fitting parameters allowed us to derive approximate expressions for the magnetization and the correlation length as functions of the temperature that provide good qualitative agreement with the dynamic Metropolis Monte Carlo simulations.

Basing on our analysis, we can conclude that:

First, computer simulations allows one to determine correctly distinctive features of behavior of spin systems even when the values of $N$ are comparatively small. Increase of $N$ only allows one to determine more accurately the values of the critical parameters. However, this refinement is not very important: from Eq. (10) we see that the accuracy of the critical values $\beta_c$ and $U_c$ is determined to within $\sim l^{-1}$.

Second, a system of a finite size has no logarithmic divergence of the heat capacity in the critical point predicted by Onsager. The same is also true for the energy dispersion. It was to be expected from the most general considerations. Instead we see that the heat capacity in the critical point increases logarithmically, that is $C_c \sim \ln N$. One would think that when $N \to \infty$ we return to the Onsager solution ($C \to \infty$ for $\beta \to \beta_c$). However, it is difficult to realize this transition since if we increase the size of the lattice even up to the Avogadro number $N \sim 10^{23}$, the heat capacity $C_c$ increases only 4 times comparing with the case $N = 10^6$. Moreover, the dependence of the heat capacity on $N$ means violation of the additivity concept for a classical system that is clearly seen even when $N \sim 10^{23}$: if we double the size of the system the value of $C_c$ is up by 1.3%. Of course, we examined the system with the free boundary conditions and such systems are non-additive by definition. In the same time, non-additivity is present also in the models with the periodic boundary conditions since according Eq. (21) in the critical point correlations increase proportional to linear sizes of the lattices.

The logarithmic dependence of the critical heat capacity on the dimension was predicted as far in the Onsager paper [3]. He examined a rectangular lattice of the size $l \times l'$ and when $l' \to \infty$ obtained the dependence:

$$C_c \approx 0.4945 \cdot \ln l + 0.1879 \qquad (24)$$

We examined a somewhat different system (a square lattice) and obtained

$$C_c \approx 0.4945 \cdot \ln l - 0.4403 \qquad (25)$$

Comparing Eqs. (24) and (25) we see that only the constant but not the dependence of $C_c$ on $l$ changed. We suppose that this is a result of different boundary conditions: Onsager used the periodic boundary conditions, while we explore the free boundary conditions. The experiment performed for a square lattice with the periodic boundary conditions [40] confirms this conclusion.

Finally, our simulations showed that the peaks of the curves $\sigma_E^2 = \sigma_E^2(\beta)$ and $C = C(\beta)$ do not coincide. At that the heat capacity reaches its maximum at larger values of $\beta$. This is an expected result since the relation between the heat capacity and the energy dispersion is $C(\beta) = \beta^2 \sigma_E^2(\beta)$. However, there is a question: which of the peaks we have to use to define the critical temperature? Indeed, the first peak corresponds to the maximum of the energy dispersion and the second to the maximum of the correlation length. They both are characteristics of a phase transitions. We (mostly through a habit) defined the critical temperature as the coordinate of the peak of the heat capacity. When the size of the lattice was large, this approach was justified: when $N \geq 400 \times 400$ the distance between these peaks was so small that we did not see it. However, at smaller sizes the distance between the peaks was noticeable. Most likely when interpreting the results of computer simulations we have to state that the phase transition is smeared over the interval from $\beta_c^*$ to $\beta_c$. Consequently, numerical experiments allow one to define the critical temperature within the accuracy of the length of this interval. This means that the absolute error has to be of the order of $\pm \beta_{ONS} / 4\sqrt{N}$.

We analyzed the dependences of the critical parameters on the size of the lattice by example of the 2D Ising model. However, we hope that our basic conclusions are also true for other models.

The work was supported by Russian Foundation for Basic Research (RFBR Project 18-07-00750).